\documentclass[notitlepage,aps,amssymb,noshowpacs,superscriptaddress,nofootinbib]{revtex4-1}	
\usepackage{graphics}
\usepackage{epstopdf}
\usepackage{enumerate}
\usepackage{pstool}
\usepackage{subfigure}
\usepackage{float}

\begin{document}
\title{Plunges in the Bombay Stock Exchange: Characteristics and Indicators}
\author{Kinjal Banerjee}
\email{kinjalb@gmail.com}
\affiliation{Department of Physics, BITS Pilani K.K Birla Goa Campus, N.H. 17B Zuarinagar, Goa 403726, India.}
\author{Chandradew Sharma}
\email{csharma@goa.bits-pilani.ac.in}
\affiliation{Department of Physics, BITS Pilani K.K Birla Goa Campus, N.H. 17B Zuarinagar, Goa 403726, India.}
\author{Bittu N}
\email{f2012514@goa.bits-pilani.ac.in}
\affiliation{Department of Physics, BITS Pilani K.K Birla Goa Campus, N.H. 17B Zuarinagar, Goa 403726, India.}
\date{\today}
\begin{abstract}

We study the various sectors of the Bombay Stock Exchange(BSE) for a period of 8 years from January 2006 - March 2014. Using the data
of daily returns of a period of eight years we investigate the financial cross-correlation co-efficients among sectors of BSE and 
Price by earning (PE) ratio of BSE Sensex. We show that the behavior of these quantities during normal periods and
during crisis is very different. We show that the PE ratio shows a particular distinctive trend in the approach to a
crash of the financial market and can therefore be used as an indicator of an impending catastrophe. We propose that a 
model of analysis of crashes in a financial market can be built using two parameters (i) the PE ratio 
(ii) the largest eigenvalue of the cross-correlation matrix(LECM). 

\end{abstract}
\maketitle

\section{Introduction}

The stock market is an extremely complex system with various interacting components \cite{Mantegna99}. 
The movement of stock prices are somewhat interdependent as well as dependent on a wide multitude of external stimuli like announcement 
of government policies, change in interest rates, changes in political scenario, announcement of quarterly results by the listed companies 
and many others. The overall result is a chaotic complex system which has so far proved very difficult to analyze and
predict. However, with so much capital invested in the financial markets, a collapse in financial markets has the potential to cause widespread 
economic disruption. As an example, the market crash of 1929 was a key factor in causing the great depression of the
1930s \cite{depression1929}.
 
It is therefore of crucial importance to be able to predict such catastrophic events so that precautions and corrective
steps can be taken. There has been an increased interest in studying and understanding the financial markets with Random
Matrix Theory proving to be an useful tool (see \cite{RMT} and references therein). It was observed that before stock market crashes, 
there was very high rise in the stock price without high increase in earning \cite{Galbraith}.  This pattern was present in almost all 
great crashes \cite{Foster}. Currently, however we do not have any accepted model characterizing the collapse of a stock market \cite{Frankel}. 
In fact there is no specific definition of what characterizes a crash \cite{predictcrash} although log-periodic power
law singularity (LPPLS) models \cite{Sornette} are promising candidates. A first step towards studying and modeling
behavior of markets before and during a crash would be to find what are the characteristics of a crash, 
i.e. to identify and study which  market parameters show distinctive behavior during the time when a financial market
collapses. 

It is well known that the market becomes very highly correlated during any period of high volatility \cite{volatility} but it is not clear
whether any such period of high correlation can be considered to be a crash. In this paper we try to analyze this question
and offer tentative proposal of a procedure to predict a crash in the financial market. To identify such market
parameters, we study the data of the daily returns of 12 sectors  of the BSE for about $2000$ days from beginning of January 2006 
to end of March 2014 \cite{BSE}. 

To understand and analyze the data we look at the logarithmic returns of the market. 
If $P_{i}(t)$ is the index of the sector $i=1,\dots,N$ at time $t$, then the \textit{(logarithmic) return} of the $i$th sector over a
time interval $ t = 1 $ to $ t= T $ days in the interval is defined as
\begin{equation}
R_{i}(t) \equiv \ln {P_{i}(t)}- \ln {P_{i}(t-1)} \label{return}
\end{equation}
For our data,  $T=1900$, the number of days we have considered, and $N =13$ because we look at the following 
12 sectors S$\&$P BSE Auto (Auto), S$\&$P BSE Bankex (Bankex), S$\&$P BSE Consumer Durables (CD), S$\&$P BSE Capital
Goods (CG) , S$\&$P BSE FMCG (FMCG), S$\&$P BSE Health care (HC), S$\&$P BSE IT (IT), S$\&$P BSE Metal (Metal), 
S$\&$P BSE Oil and Gas (Oil and Gas), S$\&$P BSE Power (Power), S$\&$P BSE Realty (Realty) 
and S$\&$P BSE Teck (Teck) and the S$\&$P BSE SENSEX (Sensex) which serves as the benchmark. We will be treating each sector as one entity in the rest
of the paper. 

Volatility is a statistical measure of the dispersion of returns for a given security or of a market index. If $R_i$ is
the return and $\mu_i$ is the return for the $i$th sector averaged over one month window respectively, we can define volatility of the $i$th
sector over a certain period of time $T_s$ as 
\begin{eqnarray}
\sigma_i(T_s) = \sqrt{\frac{1}{T_s} \sum_{t=1}^{T_s}\left( R_i(t) - \mu_i \right)^2 }
\label{volatility}
\end{eqnarray}
In the context of our data, $T_s$ will be one month. It is known that  high volatility of a market is linked to strong
correlations among it's sectors, i.e., sectors tend to behave as one during a crash \cite{volatility2}. 
In this paper we do a model independent quantitative study of the parameters which capture this feature. 

To study the correlations between the sectors we construct the monthly correlation matrix whose entries are given by
\begin{equation}
C_{ij} = \frac{\left<(R_i - \mu_i)(R_j -\mu_j) \right>}{\sqrt{(\left<R^2_i\right> - \mu^2_i)(\left<R^2_j\right> - \mu^2_j)}} \label{cormatrix}
\end{equation}
where, as before, $R_i$ is the return and $\mu_i$ is the average return for the $i$th sector respectively. The averaging is done
over a month. We will study the \textit{time evolution} of this correlation matrix, i.e. study the monthly variation of
some of the properties of the correlation matrix. Note that, unlike the analysis in \cite{ourpaper1}, 
we will mainly be focussing on periods of time when the crisis in the
market occurred and compare it with the behavior during normal times. Our goal in this paper is to identify characteristics 
during these periods which were quantitatively different from those during the the normal periods. These parameters can
be used to identify, characterize and model crashes in the future.

Apart from the behavior of correlations, in this paper we also explore the behavior of another important parameter,
the Price by Earning (PE) Ratio \cite{PEratio1}. In the period under consideration, although there have been several events of high
volatility and high fluctuations, two major events can be termed as crash {\cite{Blankenburg}}: (i) financial crisis (May 2006) and 
(ii) recession in the US (Jan 2008). We shall call periods of high volatility as \textit{crisis} periods in this paper
while the term \textit{crash} will be reserved only for the two periods mentioned above, when the global financial
markets collapsed. Some work in this direction has been undertaken in \cite{crashbehav}. The novelty in our work is that
we simultaneously study the correlations in the stock market and the PE ratio and observe that both need to be high for the
financial markets to crash.  A systematic study of both the parameters and their
behavior in the lead up to a crash has not been performed before (at least in the context of Indian markets) to the
best of our knowledge.

Our paper is organized as follows: In section (\ref{CM}) we study the properties of the correlation matrix 
and identify characteristics which were quantitatively different during the crisis periods and the normal periods. 
We show that this feature can also be captured very elegantly using the cross correlation matrix between sectors. 
In section (\ref{PE}) we study the behavior of the PE ratio before, during and after the two crashes of the stock market
mentioned above.  We study the variation of the PE ratio and show that it can be one of the most important indicators of
a crash. Further, from the analysis of the data we propose that it is possible to study, analyze and
intensity of a market crash using two very important parameters \cite{Toth}, the PE ratio and the largest eigenvalue of the correlation matrix(LECM). 
We finally end with our conclusions and observations in section (\ref{conclude}).

%%%%%%%%%%%%%%%%%%%%%%%%%%%%%%%%%%%%%%%%%%%%%%%%%%%%%%%%%%%%%%%%%%%%%%%%%%%%%%%%%%%%%%%%%%%%%%%%%%%%%%%%%%%%%%%%%%%%%%%%%%%%%%%%%%%%%%%%%%%%%%%%%%%
\section{Properties of the Correlation Matrix} \label{CM}

To understand a period of high volatility, we first need to determine which features of the correlation matrix \cite{Plerou} may be of interest during
that period. The high volatility of the financial market is related to  eigenvalues of a correlation matrix. The largest
eigenvalue of a correlation matrix (LECM) often becomes very large compared to other eigenvalues(e.g. second and third largest eigenvalues) of the correlation 
matrix during high volatility period. Hence this feature is quantified using the correlation matrix. 
In all the figures in this section, the periods of high volatility will be marked out in grey. They correspond to 
\begin{itemize}
\item May - July 2006
\item July - September 2007
\item January - March 2008
\item August - December 2008
\end{itemize}

Note that, high volatility always leads to a large value of LECM but the converse is not true. A large value of LECM
does not necessarily indicate high volatility.

\begin{figure}[ht]
		\centering\subfigure[Connectedness with $t=0.90$]{
                \includegraphics[height=3in,width=3in,angle=0]{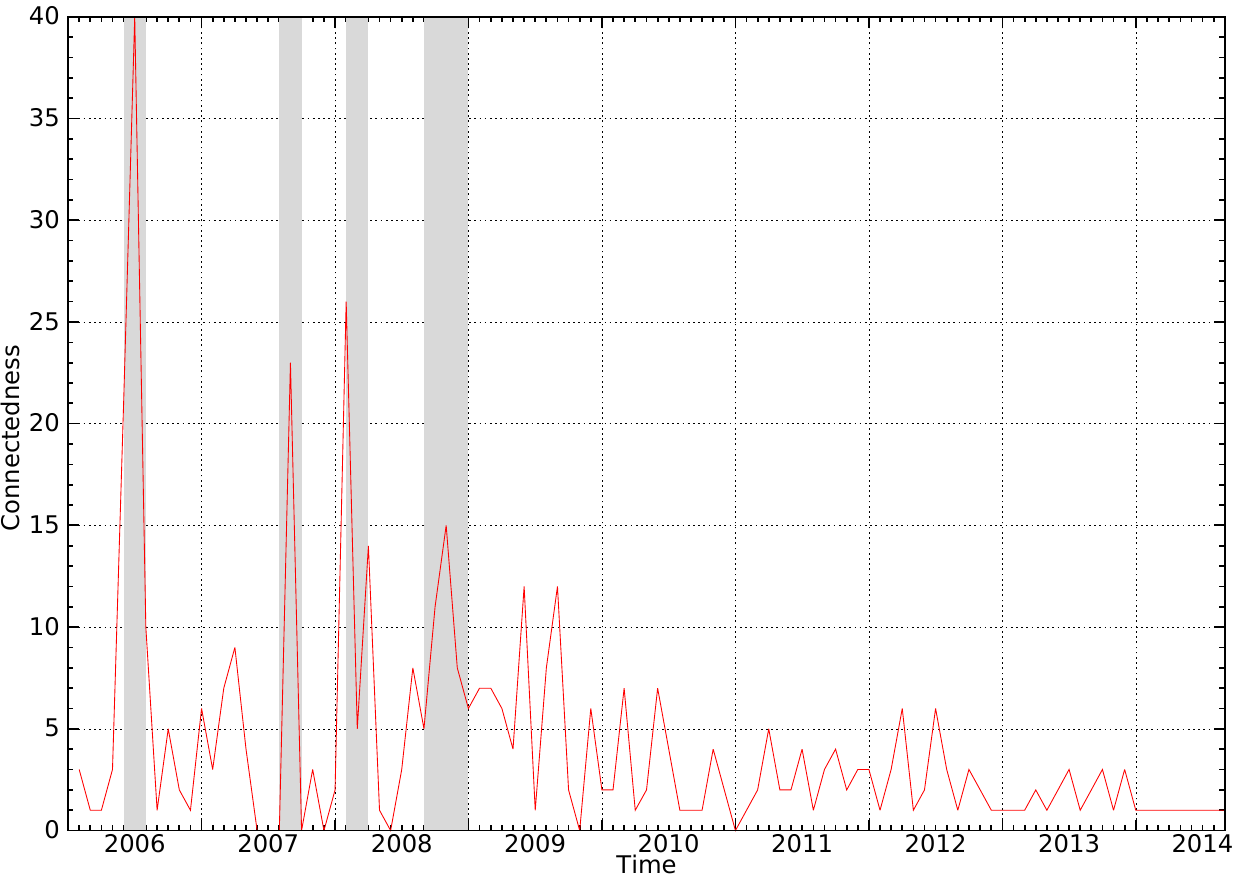}}  
\quad
\subfigure[Connectedness with $t=0.85$]{
\includegraphics[height=3in,width=3in,angle=0]{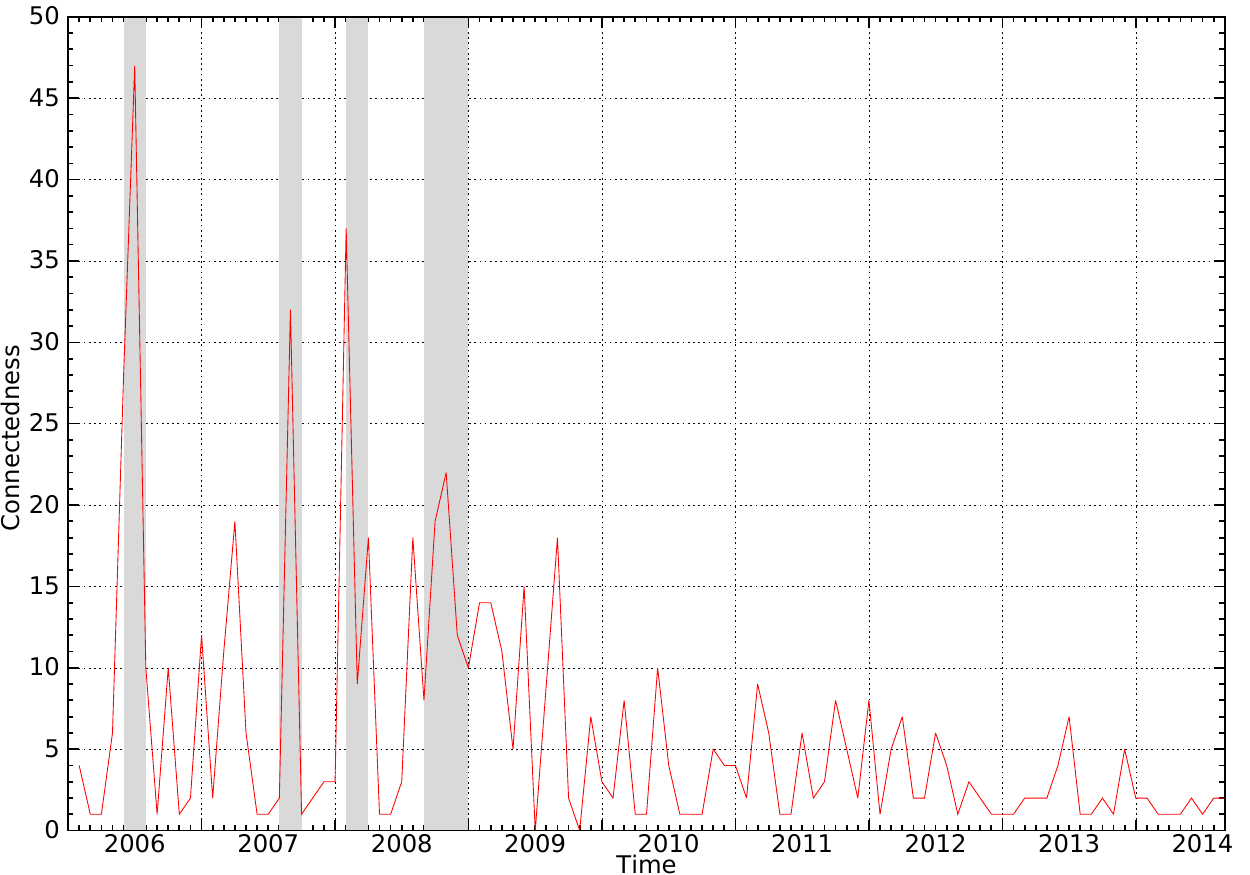}}  
		\caption{Variation of connectedness of the 12 sectors and sensex with time}
\label{connectedness13}
\end{figure}

\begin{figure}[ht]
		\centering\subfigure[Connectedness with $t=0.90$]{
                \includegraphics[height=3in,width=3in,angle=0]{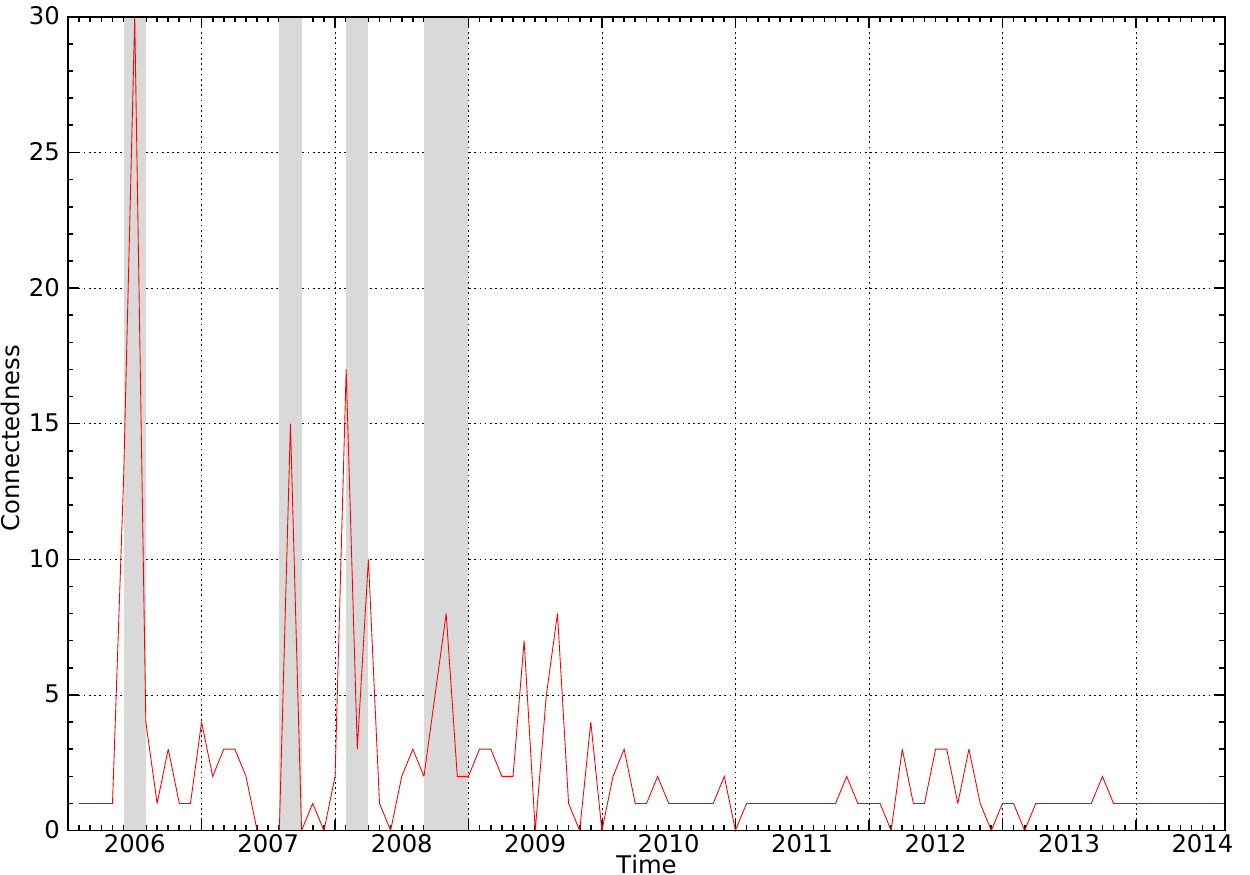}}  
\quad
\subfigure[Connectedness with $t=0.85$]{
\includegraphics[height=3in,width=3in,angle=0]{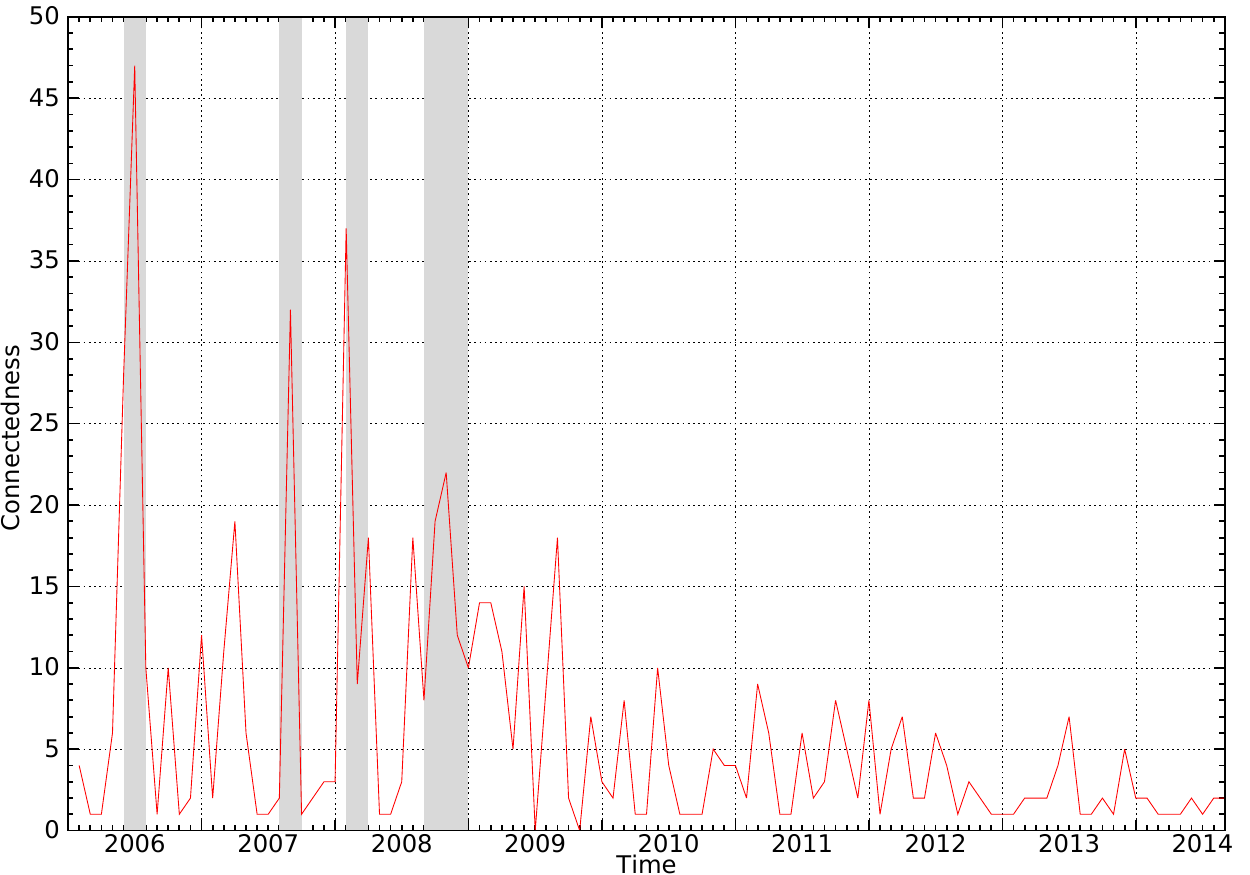}}  
		\caption{Variation of connectedness of the 12 sectors with time}
\label{connectedness12}
\end{figure}

Choosing some value of a threshold $t$, the correlation matrix is converted to the adjacency
matrix, that is, if $C_{ij} \geq t$ the corresponding entry in the adjacency matrix is $1$ and $0$ otherwise. The number of
connections i.e. the connectedness gives a measure of the amount of correlations existing in the market in a given
month. The variation of connectedness of the 12 sectors and sensex with time can be seen in
Figs.(\ref{connectedness13})and  (\ref{connectedness12}) for $t=0.9$ and for $t=0.85$. As can be seen from these two
figures, the inclusion of the sensex along with the 12 sectors does not change the variation of the connectedness
with time. Therefore in all the subsequent analysis of the features of the correlation matrix we have included the
sensex along with the 12 sectors. 

\begin{figure}[ht]
                \includegraphics[height=3in,width=6in,angle=0]{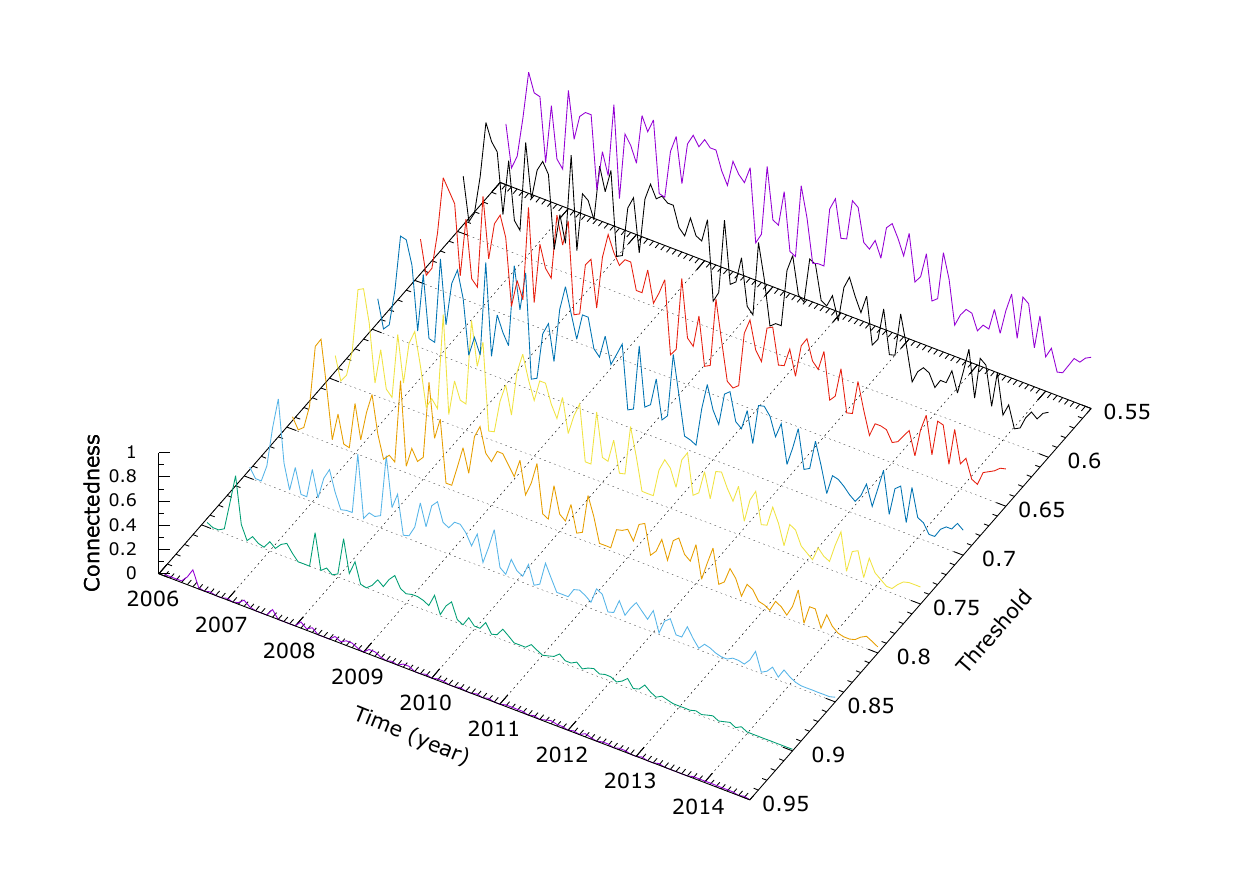} 
		\caption{(Normalized) Connectedness with $t=0.95$, $t=0.90$, $t=0.85$, $t=0.80$, $t=0.75$ and $t=0.75$}
\label{connectedness}
\end{figure}

Although the choice of threshold is arbitrary it does not change the conclusions. As can be seen
from Figs.(\ref{connectedness13}), (\ref{connectedness12}) and (\ref{connectedness}), the qualitative features do not change with changing 
$t$ but obviously the actual nature of the connectedness becomes more apparent above some value of the threshold. For convenience the value 
of $t=0.9$ can be chosen so that the number of connections during normal periods is below 10 but increases significantly during periods
of high volatility. It is also evident from these figures that correlation among sectors are quite high during crisis period and low in normal period. 
It is obvious that during periods of financial crisis (the grey regions in the figure) the market is highly correlated
and hence strongly connected.

\begin{figure}[ht]
		\centering
\subfigure[During crisis period (April 2006)]{
		\includegraphics[width=0.45\textwidth]{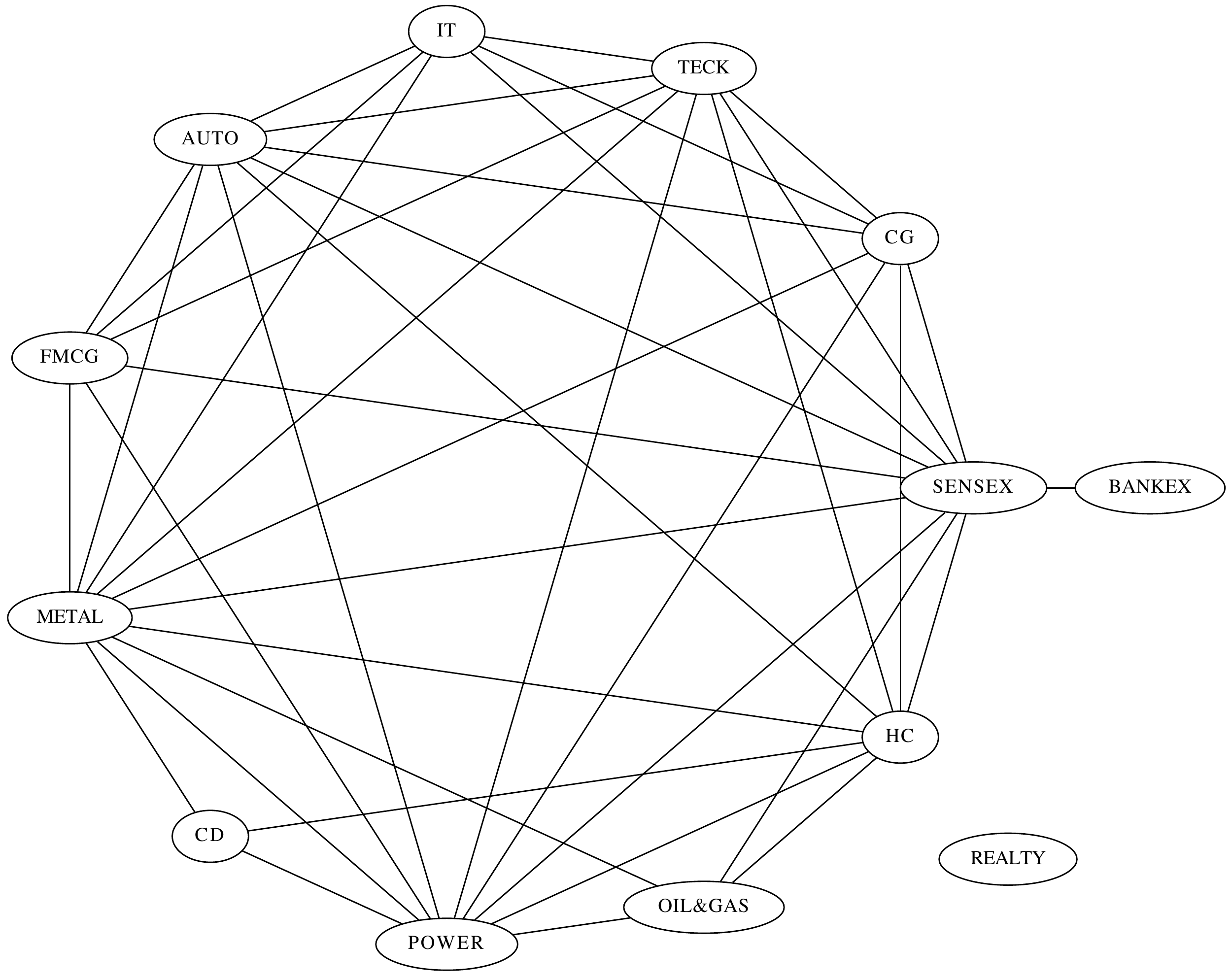}}
\quad
\subfigure[During normal period (May 2010)]{
		\includegraphics[width=0.45\textwidth]{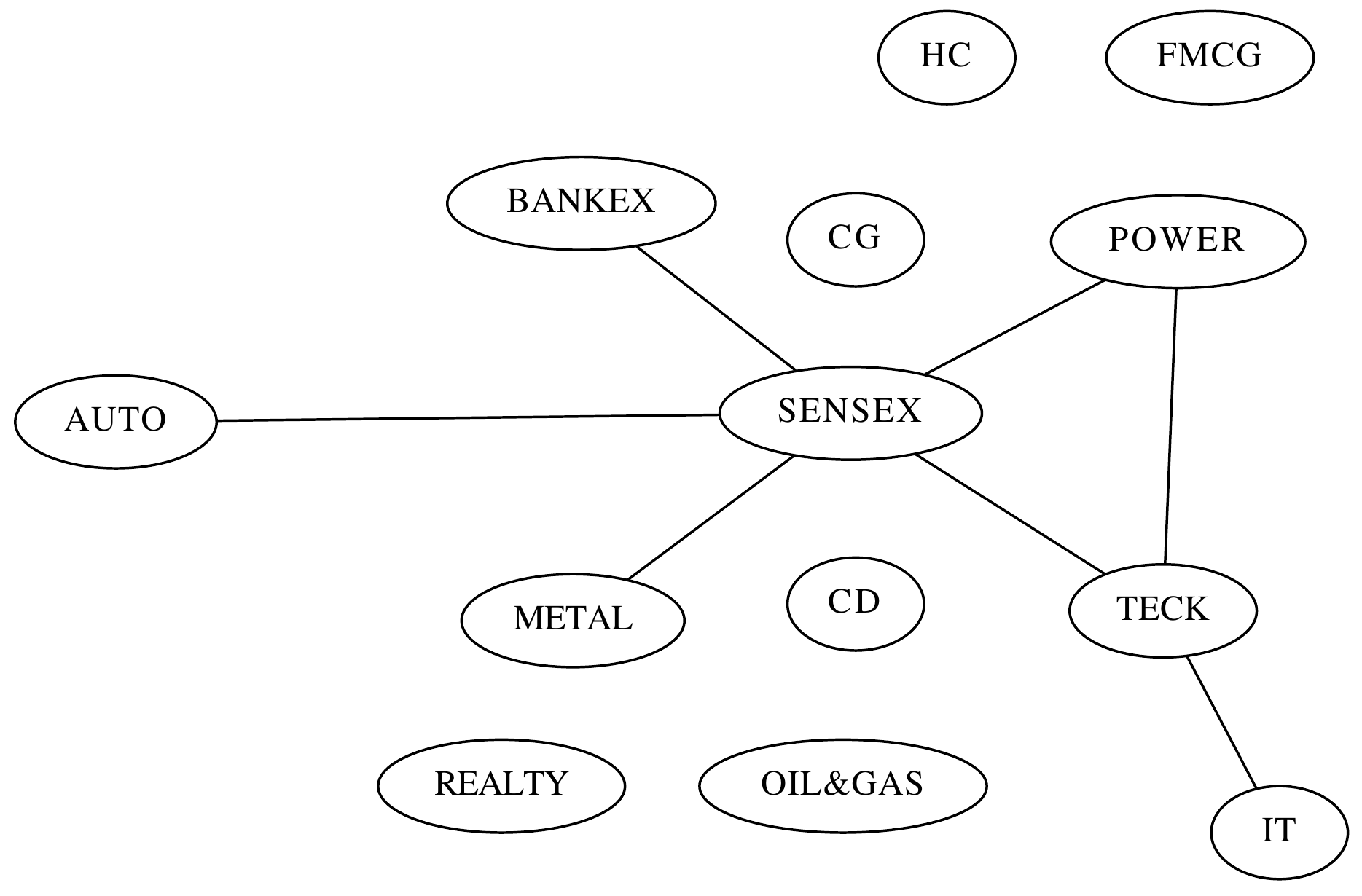}}
		\caption{Comparison of connectedness during a crisis and normal periods}
\label{connectednessgraph}
\end{figure}

We further illustrate this feature by constructing a graph from the adjacency matrix of $t=0.9$ in
Fig.(\ref{connectednessgraph}) for two representative months. In the graph, each sector is represented as a node. Two
nodes are connected by a link if the corresponding entry in the cross correlation matrix is above the threshold $t=0.9$. 
It can be clearly seen that during crisis periods (eg. April 2006) the resultant graph is highly connected with almost 
all the sectors linked to each other. That is not the case during normal periods (eg. May 2010) when the numbers of links are much less.

\begin{figure}[ht]
		\centering\subfigure[Mean value of the correlation matrix with time]{
		\includegraphics[height=3in,width=3in,angle=0]{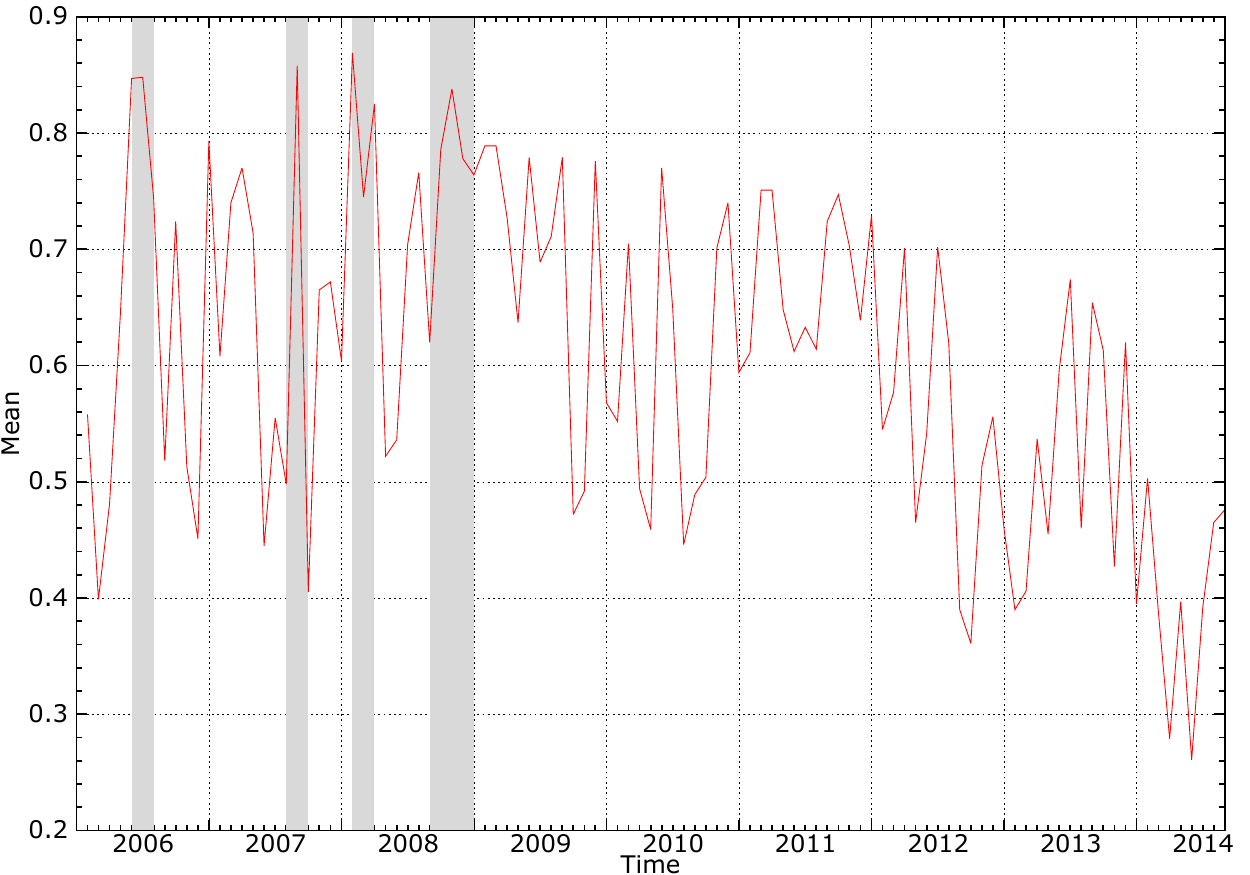}}
\quad
                \centering\subfigure[Standard Deviation of the correlation matrix with time]{
		\includegraphics[height=3in,width=3in,angle=0]{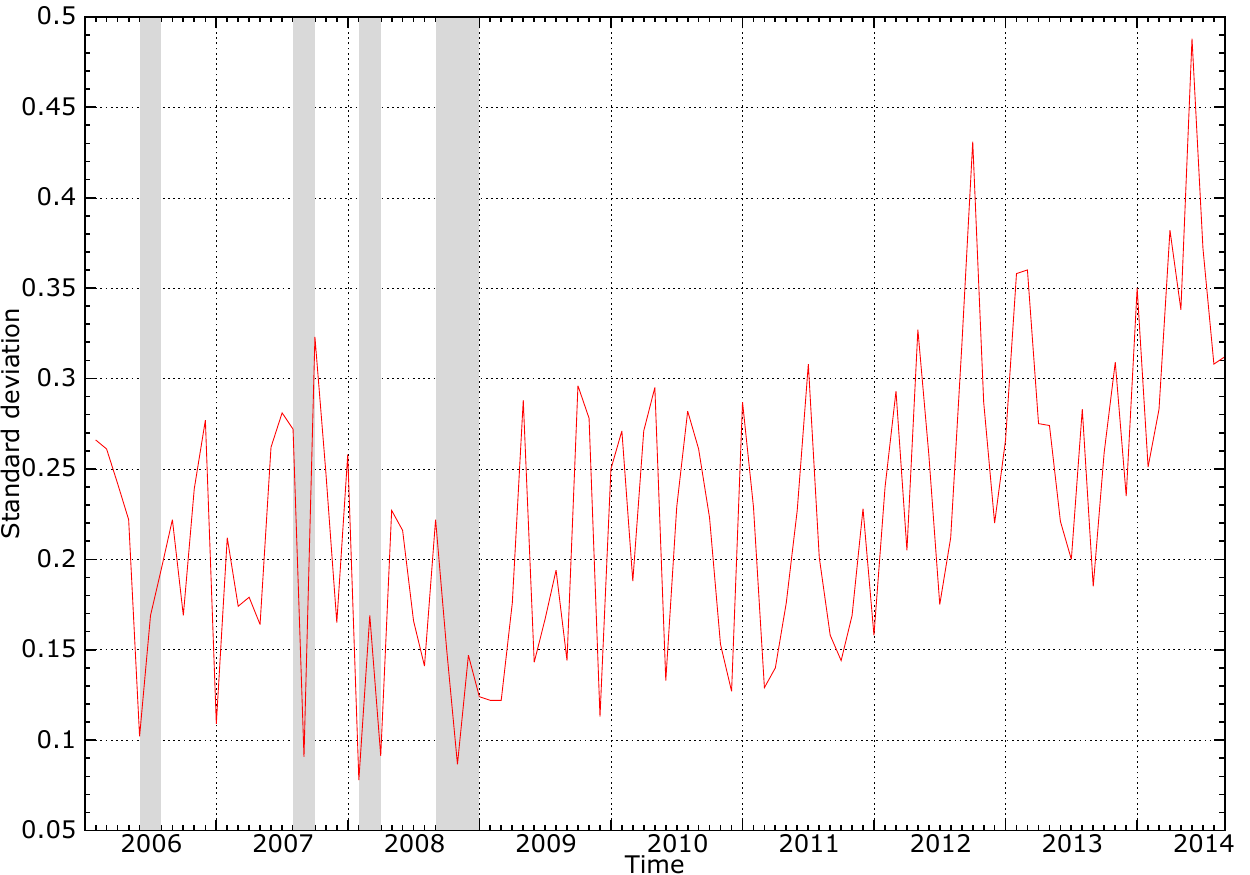}}
\quad
                \centering\subfigure[Ratio of Standard Deviation to Mean of the correlation matrix with time]{
		\includegraphics[height=3in,width=3in,angle=0]{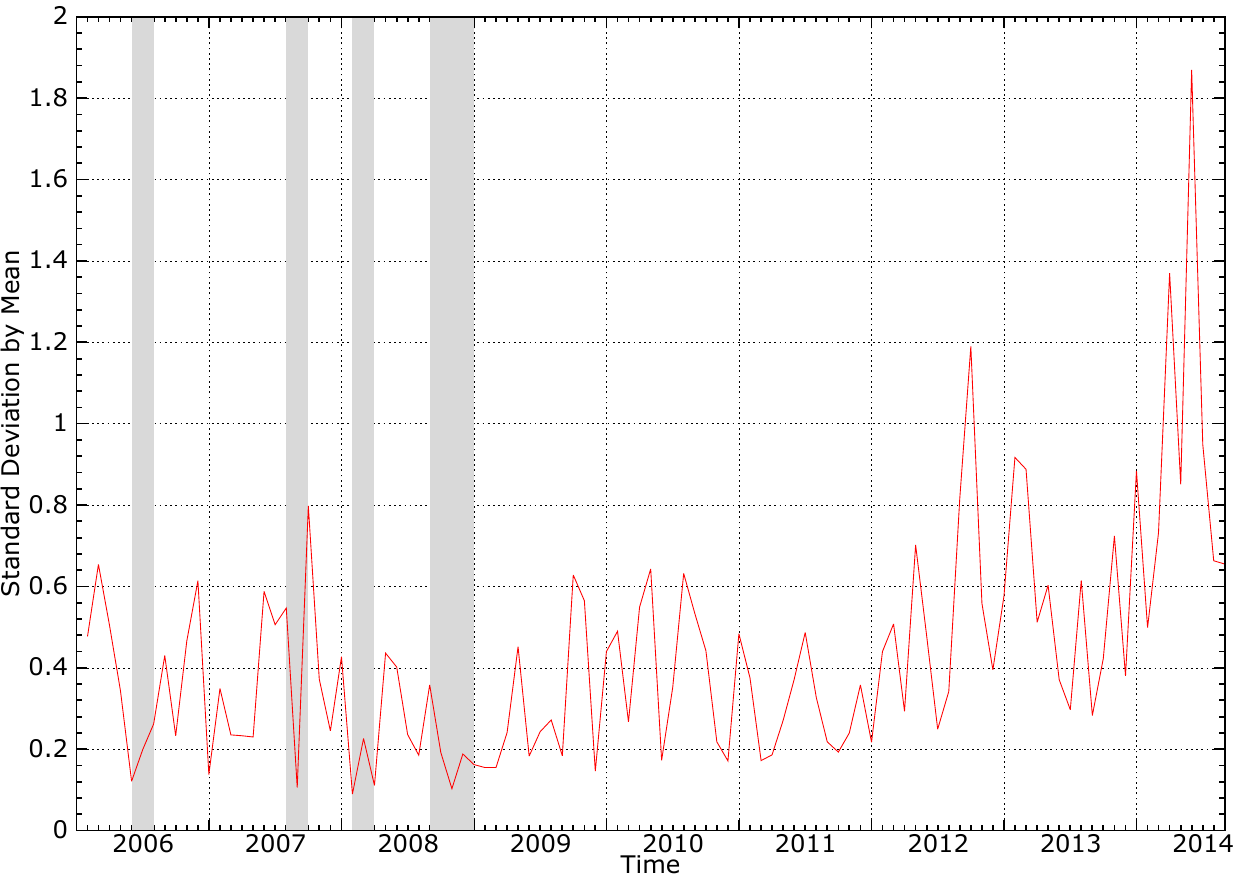}}
    \caption{Variation of Mean, Stdev. and Stdev.to Mean with time}
\label{meanstd}
\end{figure}

Another way to characterize periods of crisis is to look at the mean, standard deviation and the ratio of standard deviation to mean of the correlation matrix in Fig.(\ref{meanstd}).
Periods of high mean correlation and low standard deviation signify a highly correlated market \cite{volatility3}. 
During normal periods the mean is around or below 0.80 and during a crisis, the value is above 0.80 such as in April 2006 when it touched 0.85. 
Similarly, during normal periods the standard deviation is around or above 0.12 and during a crisis, 
the value is much lower such as in April 2006 when it dropped to 0.10. Similar features can be seen in the ratio of
standard deviation and mean.

\begin{figure}[ht]
		\centering\subfigure[Smallest element of the correlation matrix]{
		\includegraphics[width=0.65\textwidth]{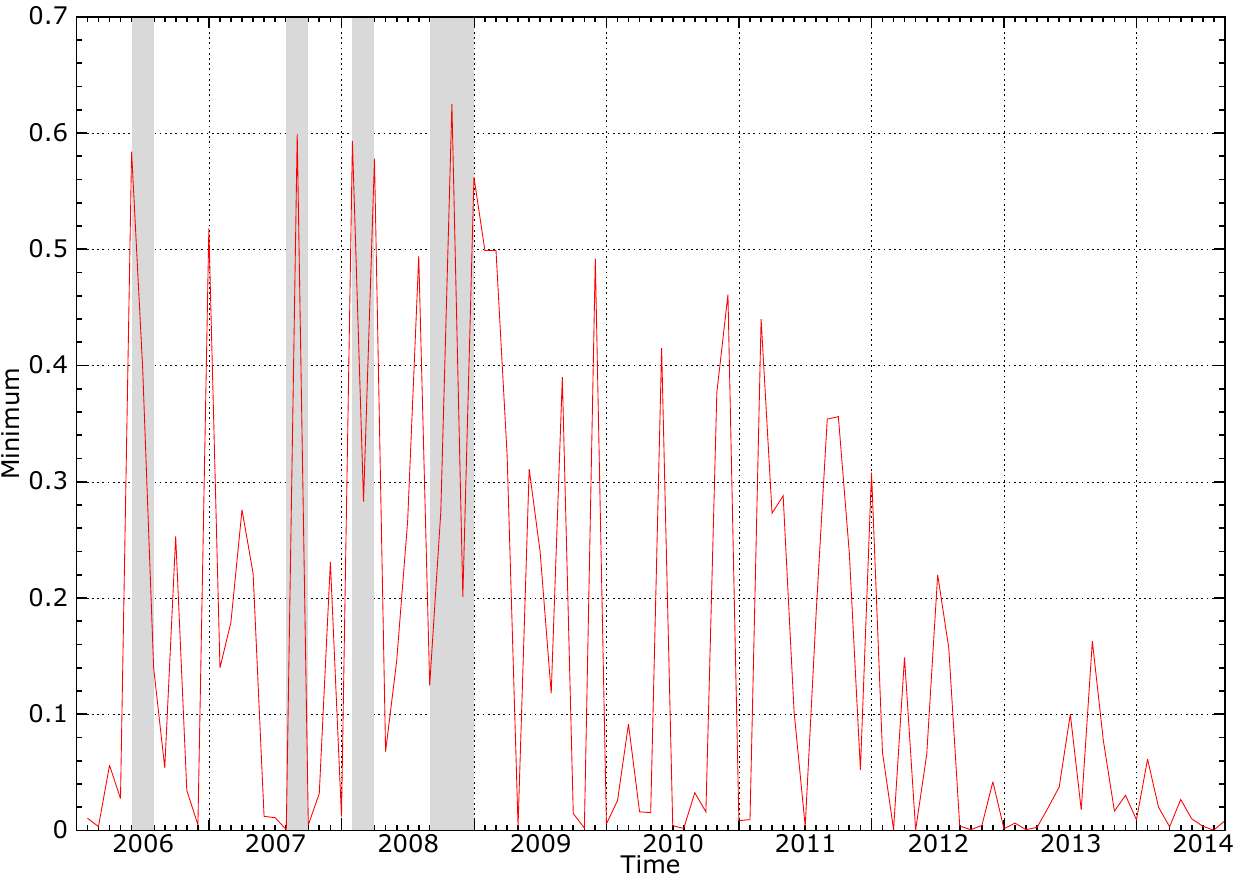}}
    \caption{Variation of Minimum Correlation}
\label{maxmin}
\end{figure}

Some of the individual entries of the correlation matrix also contain information about large volatility \cite{Iori}. 
A relatively large value of minimum correlation suggests that sectors that would otherwise not be related, are moving together. 
As can be seen in  Fig.(\ref{maxmin}) during normal periods its value is a way  below 0.65 and during a crisis the value is above 0.65. 

\begin{figure}[ht]
	\centering
	\includegraphics[width=0.65\textwidth]{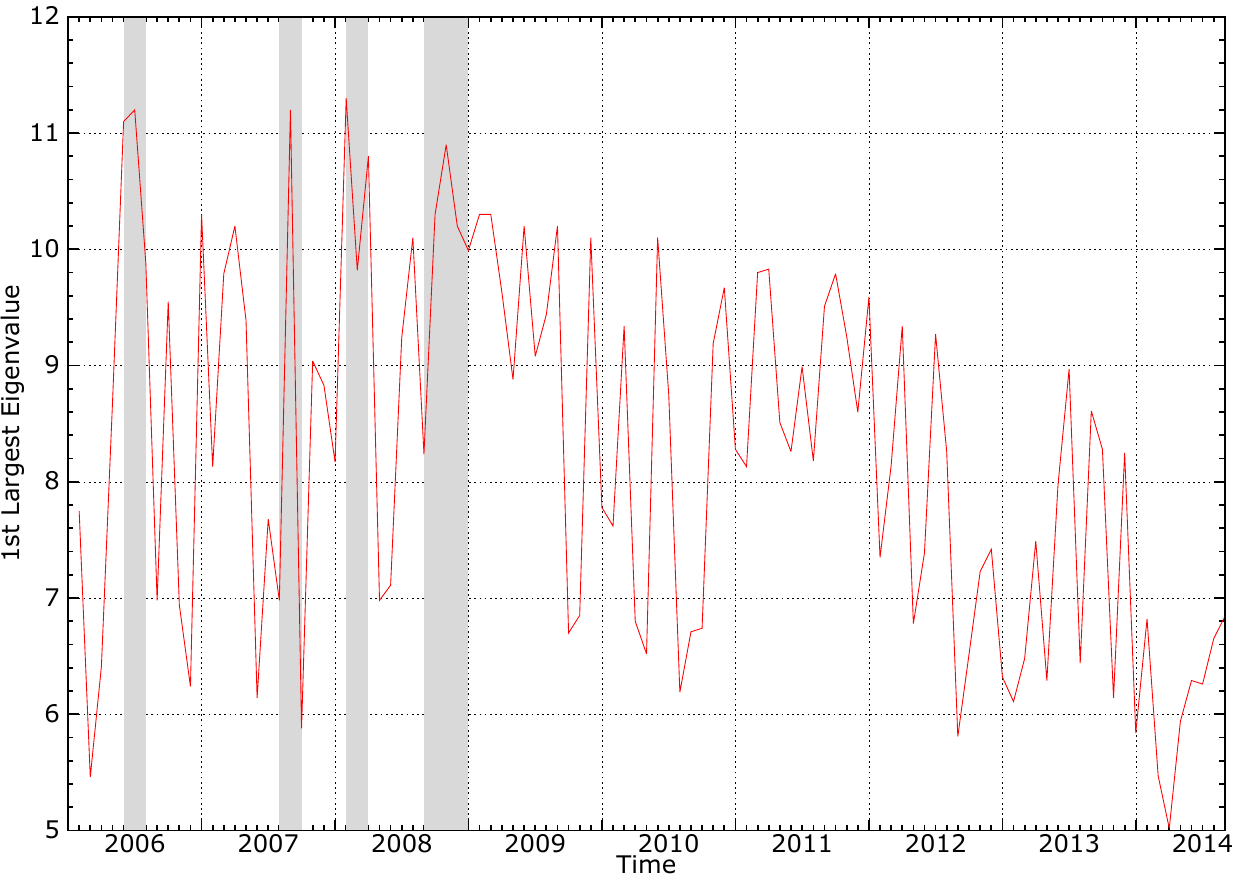}
	\caption{Largest eigenvalue of the correlation matrix}
\label{maxev}
\end{figure}

\begin{figure}[ht]
	\centering
\subfigure[Second Largest Eigenvalue]{
	\includegraphics[width=0.45\textwidth]{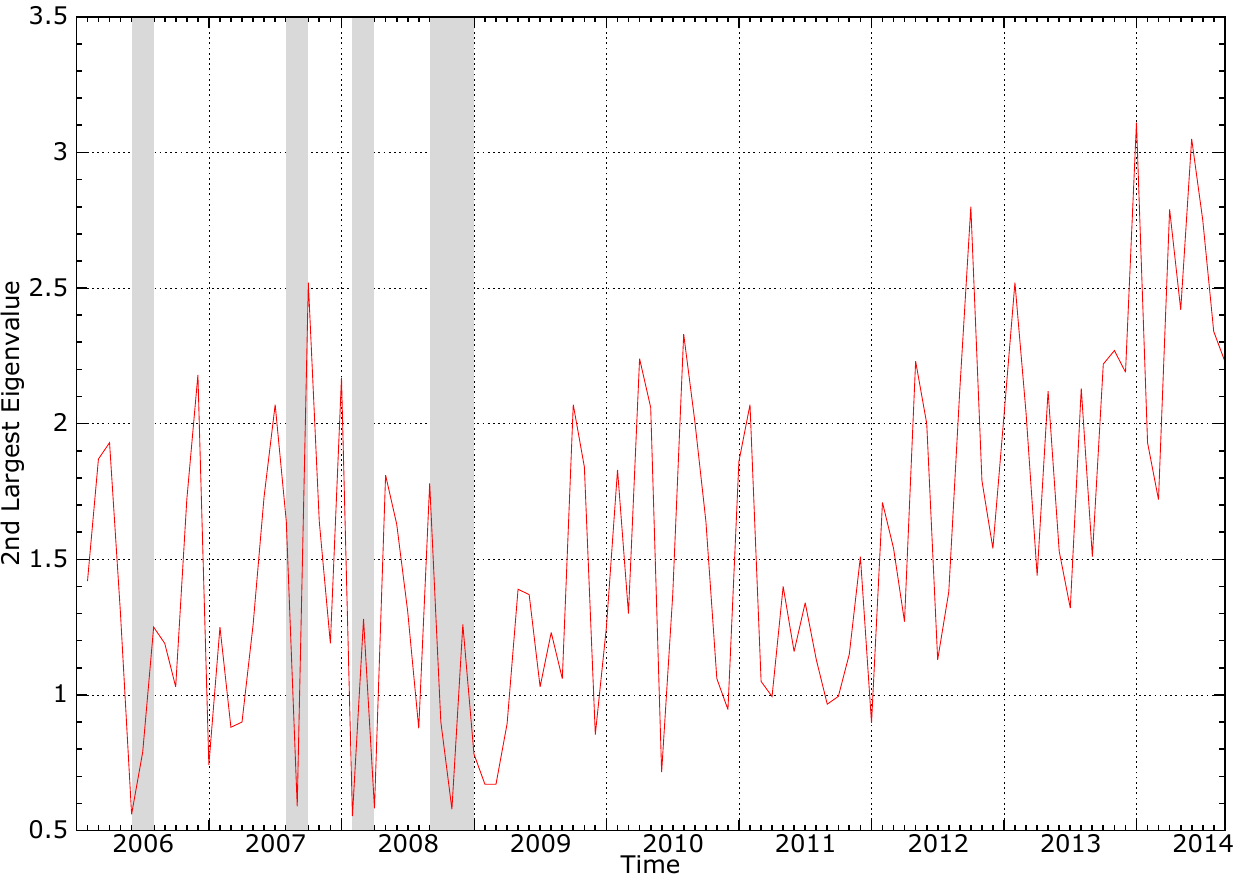}}
\quad
\subfigure[Third Largest eigenvalue]{
        \includegraphics[width=0.45\textwidth]{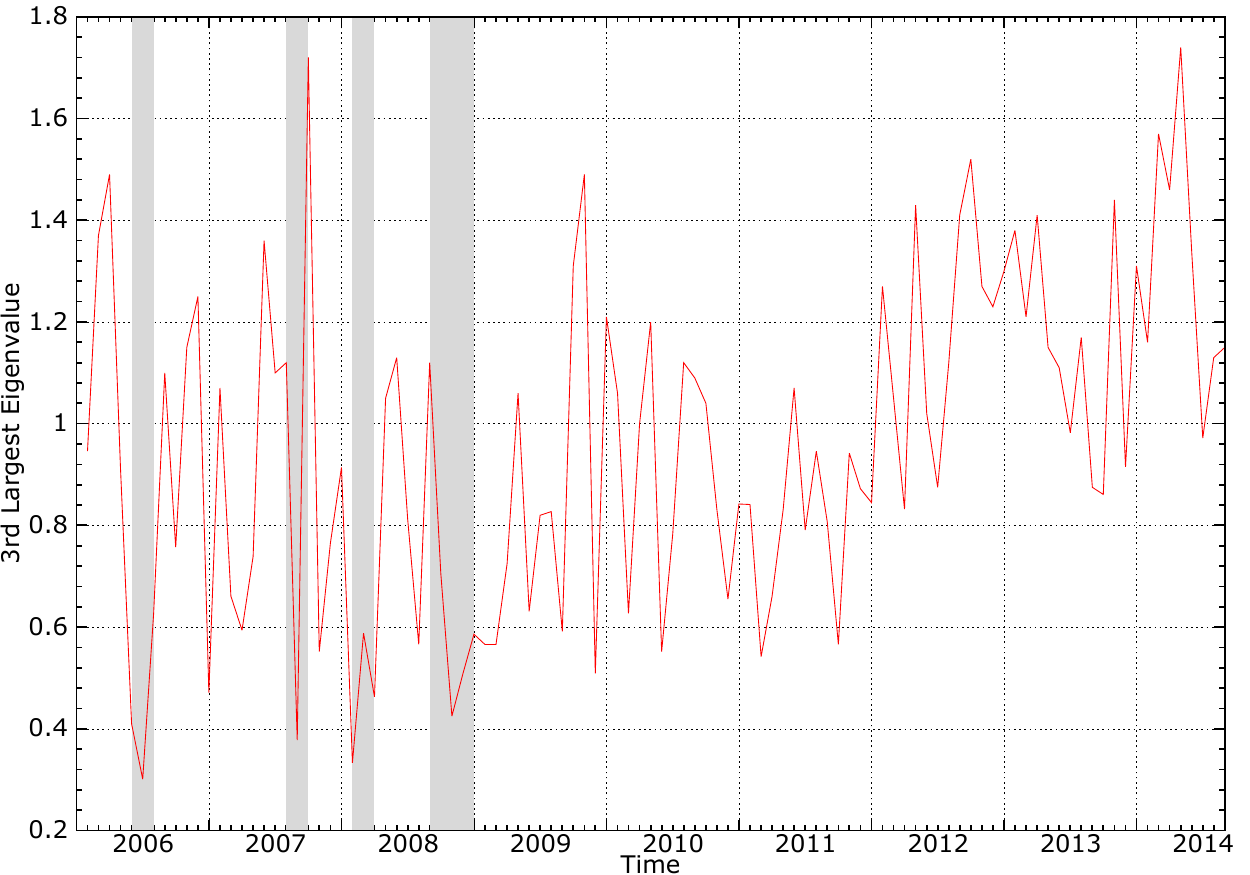}}
	\caption{Second and Third largest eigenvalue of the correlation matrix}
\label{maxev23}
\end{figure}

Another good measure of the correlation existing in the market can be obtained from studying the largest eigenvalue of
the correlation matrix of the 12 sectors and sensex. We shall denote the largest eigenvalue of the correlation matrix as
LECM subsequently. 
The largest eigenvalue of a correlation matrix indicates the maximum amount of the variance of 
the variables which can be accounted for with a linear model by a single underlying factor \cite{Friedman1981}. 
A large value is an indicator of a common driving force behind the entire market \cite{volatility4}. This is especially relevant during a
crisis. From Figs.(\ref{maxev}), we can see that during normal periods the largest eigenvalue is  below 10 and 
during a crisis the value is closer to 11. We can also see clearly from Fig.(\ref{maxev23}) that second and third largest 
eigenvalues of the correlation matrix of 12 sectors and sensex are low during crisis periods but are quite high during
normal periods.

Note that, as expected \cite{Friedman1981}, the graph for the mean value of the correlation matrix (Fig.(\ref{meanstd}a)) and that of the
largest eigenvalue of the correlation matrix (Fig.(\ref{maxev}) is qualitatively similar. However we choose to work
with largest eigenvalue instead of the mean value of correlations because it is more connected to Random Matrix Theory.
Subsequently, in this paper, we use LECM to capture the presence of high correlations in the financial market. 

Therefore we can argue that any crisis period of a stock market can therefore be characterized by the following properties
\begin{itemize}
\item high volatility
\item high connectedness among sectors
\item low standard deviation of correlations
\item high mean correlation
\item high maximum correlation
\end{itemize}

In all the figures above, the grey regions denote periods of high volatility. These periods can be checked against
historical data of SENSEX ( see Table (\ref{historic})) to justify our assertion from the table.
\begin{table}[ht]\label{historic}
\parbox{1 \linewidth}{
\centering
\begin{tabular}{| l | l | l |l|}
 \hline
    \bf{Period} & \bf{Closing Index (min)} &  \bf{Closing Index (max)} & \bf{Remark}\\ \hline
    May-July 2006 & 8929.44 & 12612.38 & three consecutive large fall (May 18, 19, 22) \\  \hline
    July- Sept 2007 & 13989.11 & 17291.1 & no consecutive large fall\\ \hline
    Jan- March 2008 & 14809.49 & 20873.33 & three consecutive large fall (Jan 18, 21, 22)\\ \hline
    Aug-Dec 2008 & 8451.01 & 15503.92 & large fall but no consecutive  fall\\ \hline
\end{tabular}
\caption{Historical Data (SENSEX)}
}
\end{table}

Note that our analysis is done with monthly averages and consequently our results will only indicate high volatility
periods which lasted at least more than a month.   

%%%%%%%%%%%%%%%%%%%%%%%%%%%%%%%%%%%%%%%%%%%%%%%%%%%%%%%%%%%%%%%%%%%%%%%%%%%%%%%%%%%%%%%%%%%%%%%%%%%%%%%%%%%%%%%%%%%%%%%%%%%%%%%%%%%%%%%%%%%%%%%%%%%%%%%%%%%%%%%%

\section{Major Plunges in BSE} \label{PE}

\begin{figure}[ht]
		{\includegraphics[width=0.65\textwidth]{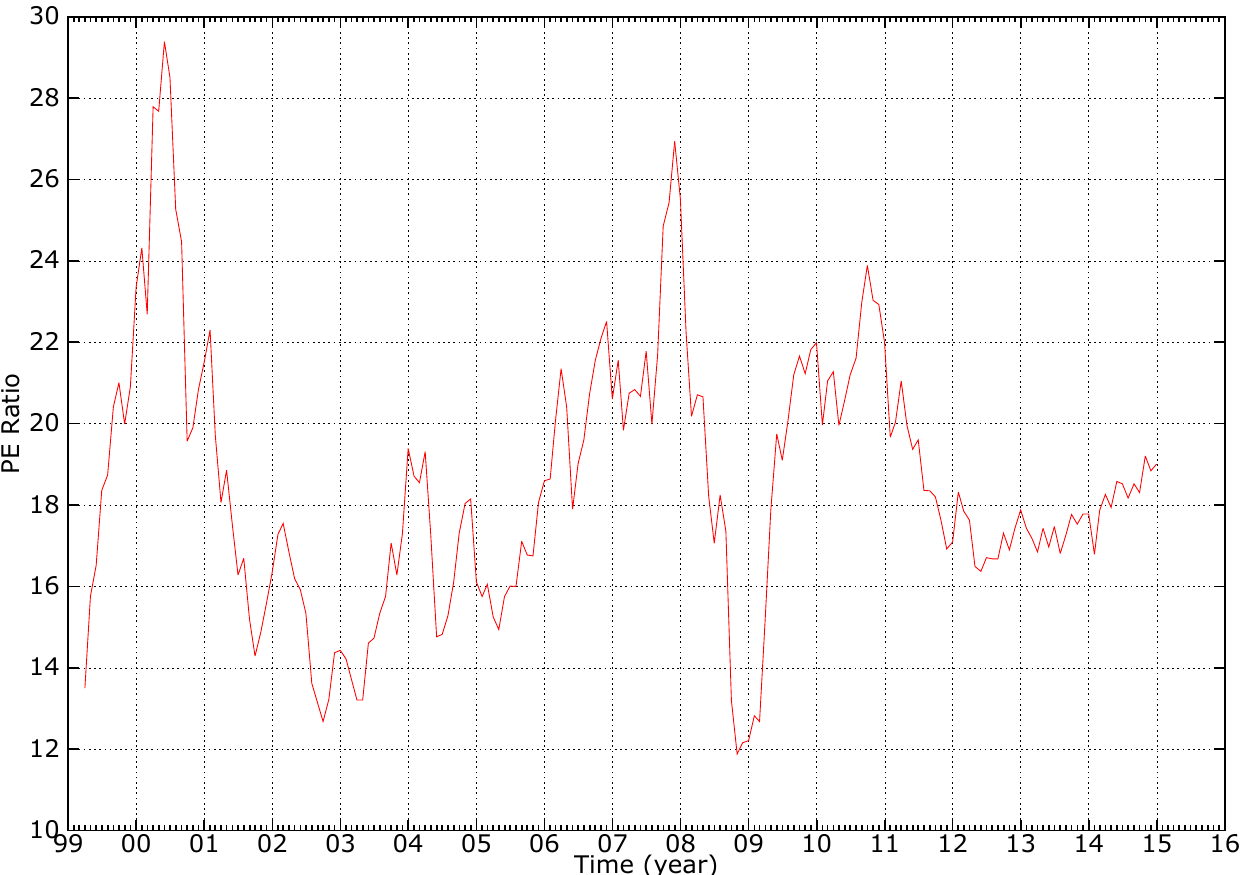}}
    \caption{Variation of PE Ratio over time}
\label{PEratio}
\end{figure}

\begin{figure}[ht]
		\centering\subfigure[Variation of LECM and PE ratio with time]{
                \includegraphics[width=0.45\textwidth]{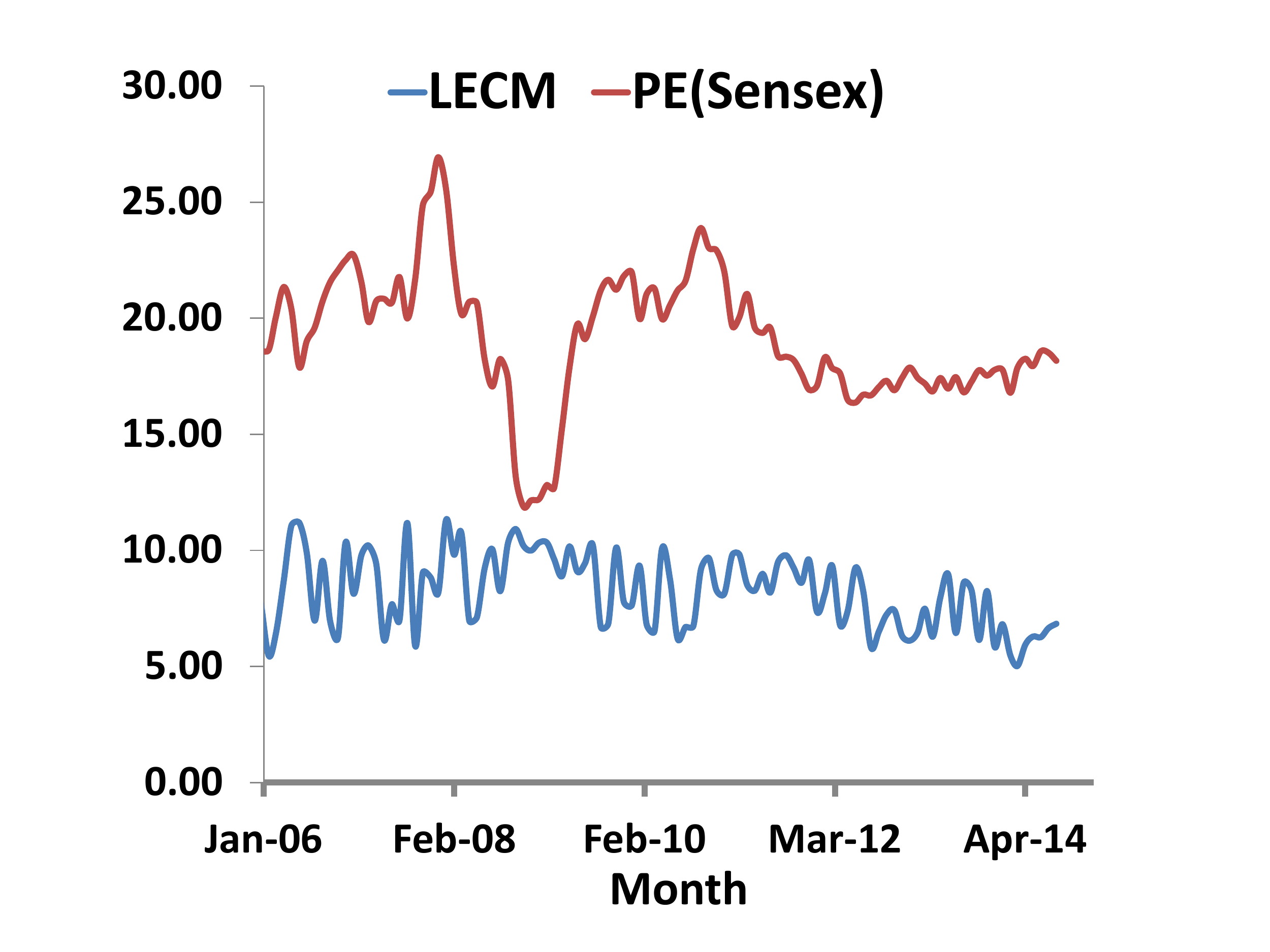}}  
\quad
\subfigure[Parameter Space of LECM and PE ratio]{
\includegraphics[width=0.45\textwidth]{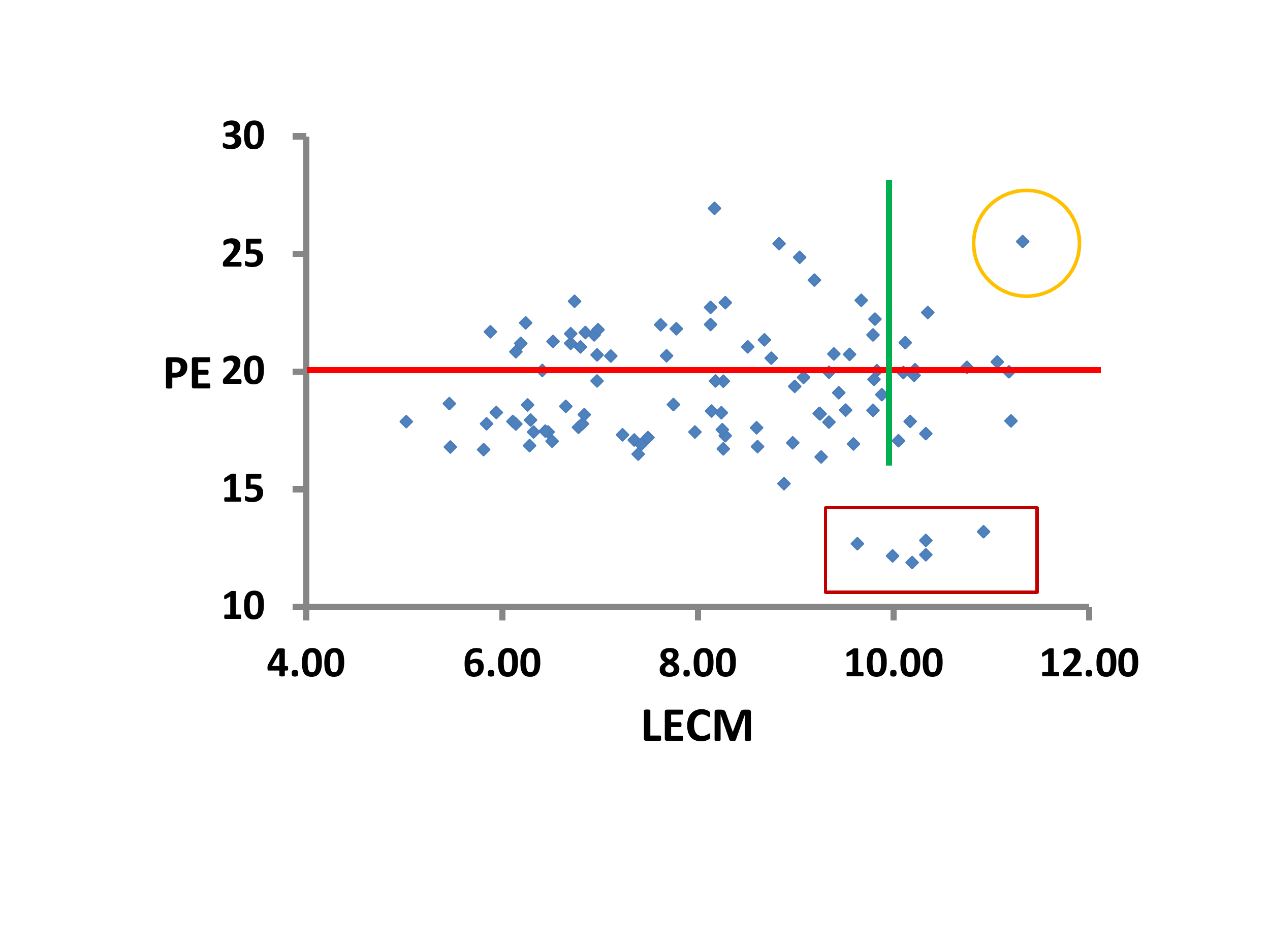}}  
		\caption{LECM and the PE ratio}
\label{LECMPE2d}
\end{figure}

In this section, we will explore the behavior of the PE ratio(PE) for the period under consideration. Let us first
qualitatively explain why we expect the PE ratio to be a significant indicator of a upcoming crisis. 
Most of the crashes of the financial markets are part of the boom and bust cycle which is captured by the PE ratio
\cite{PEratio1,PEratio2}. 
Note that, earnings, book value and dividend for each stock is assumed to be constant for each quarter 
and only change when the next quarterly results are announced. 
In the absence of any external news the price fluctuation of a stock will be almost random over a quarter. If however the market sentiment is positive
and external factors indicate that the next quarter results will show increased earning, the movement trend of the
stock price will be in the upward direction. If this sentiment continues for longer time, the rate of increase of 
the stock price will rise faster than the rate of increase based on the expected increase of earnings. Owing to
correlations present in the market, this will cause the upward movement of other stocks thereby increasing the PE ratio. 
If the positive sentiment sustains for  longer  period, the PE ratio will become very high. Consequently the market will become unsustainable. Note that, during this
period the market movement cannot be explained on the basis of the quarterly results on earnings and book values. 
A continuous stream of external \textit{good news} affect the market and there are more buyers than sellers. The imbalance due to consistent 
good news  drives the market to new high for some times. This state of the market is a very unstable equilibrium which
can be disturbed by a stream of \textit{bad news}. Thereafter the market crashes and normalcy is established
again \cite{PEratio3}. In Fig. (\ref{PEratio}) we plot the variation of the PE ratio over time while Fig
(\ref{LECMPE2d}.a) gives the variation of both the LECM (of the 12 sectors and the sensex) and the PE ratio over time. 

Therefore the PE ratio can be used as a measure of the unsustainability of the market 
while the LECM can be used to measure how much the market as a whole is susceptible to external news \cite{Lilo}. 
For a given (high) value of LECM and same intensity of negative external news, a market with very high PE ratio 
is more likely to crash than the market with low PE ratio.
The analysis of LECM was done in the previous section (see Fig. (\ref{maxev})). 
From historical data we know that there have been two major crashes in the BSE in this time period (i) May 2006 (ii) January 2008. A consistently high monthly PE ratio of sensex 
(25 $\pm $ 5) and high LECM ($\geq 11$) is a signature of highly non-equilibrium market leading to a crash. 
In Tables (\ref{table1},  \ref{table2} )
we see the variation of the PE ratio and the largest eigenvalue of the correlation matrix (LECM) before, during and after these
two events.

\begin{table}[ht]
\parbox{.45\linewidth}{
\centering
\begin{tabular}{| l | l | l |}
 \hline
    \bf{Month} & \bf{LECM} &  \bf{PE Ratio of Sensex} \\ \hline
    Jan 06 & 7.75 & 18.6 \\ \hline
    Feb 06 & 5.46 & 18.64 \\\hline
    Mar 06 & 6.4 & 20.04 \\\hline
    Apr 06 & 8.68 & 21.35 \\\hline
    \bf{May 06} & \bf{11.05} & \bf{20.41} \\\hline
    Jun 06 & 11.2 & 17.9 \\\hline
    \hline
\end{tabular}
\caption{Around May 2006}
\label{table1}
}
\hfill
\parbox{.45\linewidth}{
\centering
\begin{tabular}{| l | l | l |}
\hline
    \bf{Month} & \bf{LECM} &  \bf{PE Ratio of Sensex} \\ \hline
    Oct 07 & 9.04 & 24.86 \\\hline
    Nov 07 & 8.83 & 25.44 \\\hline
    Dec 07 & 8.17 & 26.94 \\ \hline
    \bf{Jan 08} & \bf{11.32} & \bf{25.53} \\ \hline
    Feb 08 & 9.82 & 22.23 \\ \hline
    Mar 08 & 10.76 & 20.18 \\ \hline
    Apr 08 & 6.98 & 20.71 \\ \hline
    May 08 & 7.11 & 20.66 \\ \hline
    Jun 08 & 9.25 & 18.22 \\\hline
\hline
\end{tabular}
\caption{Around Jan 2008}
\label{table2}
}
\end{table}

\begin{figure}[ht]
		\centering
		\includegraphics[width=1\textwidth]{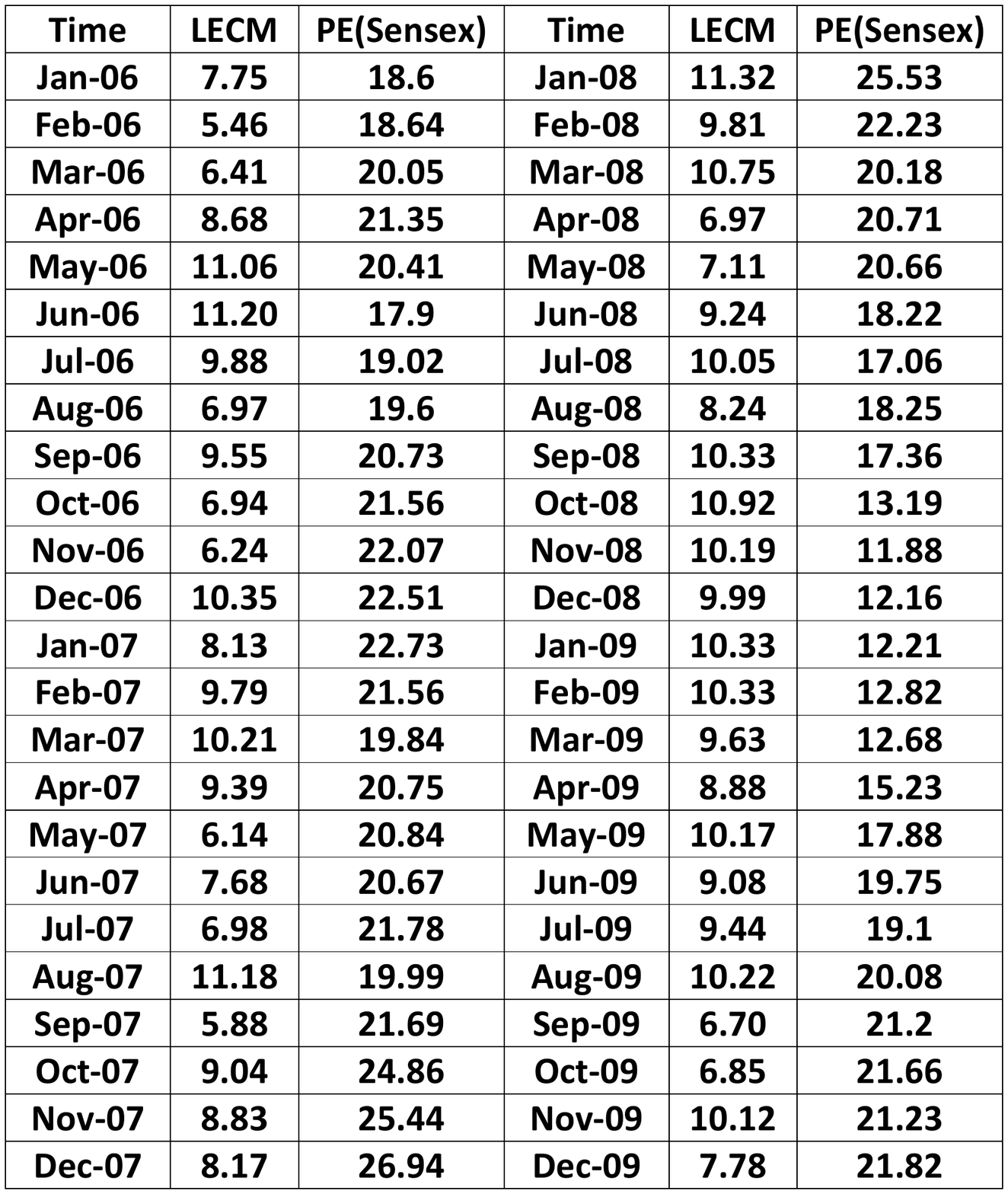}
    \caption{ Variation of  LECM of 12 sectors with sensex and PE(Sensex) for period Jan 2006-Dec 2009}
\label{tableno1}
\end{figure}

\begin{figure}[ht]
		\centering
		\includegraphics[width=1\textwidth]{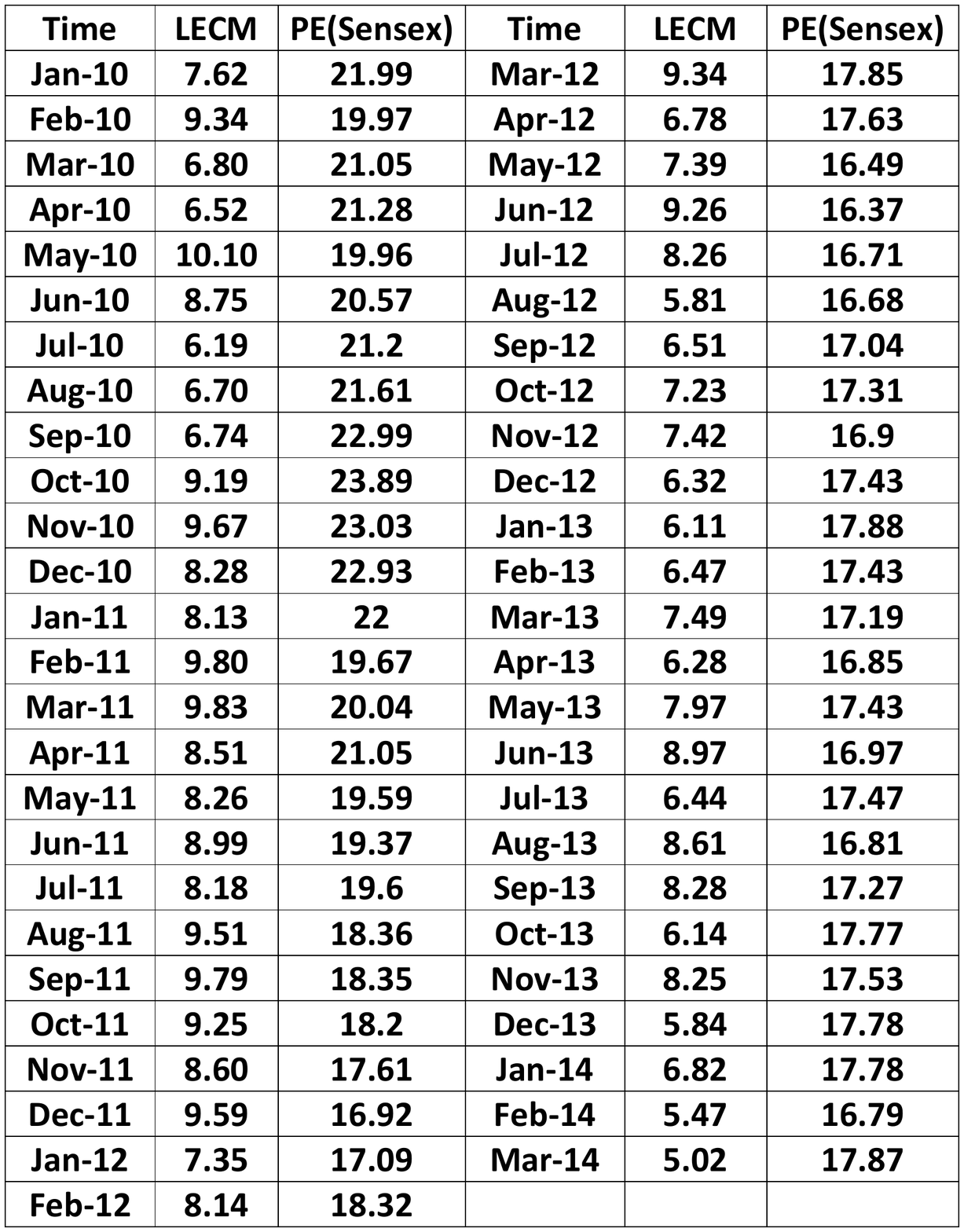}
    \caption{Variation of  LECM of 12 sectors with sensex and PE(Sensex) for period Jan 2010-Apr 2014}
\label{tableno2}
\end{figure}

A high value of PE ratio indicates an unsustainable market while a high LECM indicates a higher likelihood for a
fluctuation in one sector to affect the entire market. From the Fig. (\ref{LECMPE2d}) and the data in Tables
(\ref{table1}, \ref{table2}) and Figures(\ref{tableno1} and  \ref{tableno2} ) 
we can make several comments:
\begin{itemize}
\item The PE ratio as well as LECM at the time of crash in May 2006 was lower than that in January 2008. Consequently the
severity of the crash was lower in 2006.
\item Unlike May 2006, the PE ratio of the Sensex was consistently higher for several months before the January 2008
crash. We can conclude that before 2006 crisis, the market was slightly away from normal, while before 2008 crisis 
the market was at very high non-equilibrium  state.
\item From Fig. (\ref{PEratio}) we can see that there were other occasions when the PE ratio was above 20 but from
Fig. (\ref{maxev}) we can see that LECM was low during those times and hence the market did not crash.
\item In Fig. (\ref{LECMPE2d}.b) we have plotted the parameter space of LECM and PE ratio. The data point marked in yellow
circle is January 2008 when both LECM and PE ratio were high. That was the onset of the crisis and the market went free fall till
November 2008. This is shown by the values in the red box.
\end{itemize}
Our conclusion can also be easily verified by looking at Figs (\ref{tableno1} and \ref{tableno2}) which lists out the PE
ratio and LECM of the entire period of time under consideration.

To sum up, a large value of LECM  means that almost all sectors are moving in the same direction. It seems natural that 
almost all sectors move in the same direction during large plunge period. This implies that we should have large principle 
eigenvalue during crisis period. This is confirmed by our analysis. Our study also shows that the converse is not true; 
we may get large principle eigenvalue sometimes during normal/high volatility periods. As the data shows, a large PE ratio is required
along with a high LECM to push the market towards crash. In this paper, we show that a simultaneous study of LECM and PE
ratio is a good indicator of a crash in the financial market. In fact, these may be used to also forecast a crash in the
near future but the severity of the crash would depend on the prevailing market conditions \cite{Zhou}). 
Moreover the duration of the crash depends on how long the negative sentiment 
persists in the market, which, in turn, depends on the duration of the negative news.

%%%%%%%%%%%%%%%%%%%%%%%%%%%%%%%%%%%%%%%%%%%%%%%%%%%%%%%%%%%%%%%%%%%%%%%%%%%%%%%%%%%%%%%%%%%%%%%%%%%%%%%%%%%%%%%%%%%%%%%%%%%%%%

\section{Conclusions} \label{conclude}

The goal of this paper was to find indicators of a collapse of a stock market from studying the daily returns of BSE for
8 years. Let us emphasize once again that since our data set is the daily returns, all our statements pertain to periods
of high volatility which lasted longer than a month at least. In particular, we will not be able to detect single day
plunges of the stock market if it did not lead to (or was a part of) a sustained period of large volatility. With that
caveat in mind let us briefly recapitulate what we have done. 

It is well known that when unrealistic expectations of returns pushes the stock prices to extremely high values, the
market corrects itself through a period of collapse. It is also well known that during a collapse of the market, the 
indices of all sectors  go down irrespective of whether they were performing well or not. The first feature can be
characterized by a high PE ratio while the second feature can be characterized by a high value of the largest eigenvalue
of the cross correlation matrix (LECM). This indicates that studying the behavior of both these parameters, PE ratio and
LECM, together may be a way of characterizing a crash.  

That is exactly what we carry out in this paper. We show that by simultaneously studying the behavior of the PE ratio
and LECM, we may actually determine the times when the crashes actually occurred. We also suggest that a consistent high
value of the PE ratio along with a high LECM is a very strong indicator of an unstable market which is facing an
impending collapse. One of the major advantages of our result is that it requires only two parameters for the
prediction. This is expected to be of significant interest from the perspectives of portfolio management and economic
policy decisions. It may also be of great value in future modelling of financial markets. However, this study has to be extended 
to other markets and other economies to strengthen the result. We also expect that the volume of shares traded may play
a crucial role in studying crashes. This aspect, as well as a qualitative and quantitative study of some other market 
along the lines of the analysis carried out here is a direction of future work.

%%%%%%%%%%%%%%%%%%%%%%%%%%%%%%%%%%%%%%%%%%%%%%%%%%%%%%%%%%%%%%%%%%%%%%%%%%%%%%%%%%%%%%%%%%%%%%%%%%%%%%%%%%%%%%%%%%%%%%%%%%%%%%%%%%%%%%%%%%%%%%%%%%%%

\end{document}